\documentclass[11pt]{article}
\usepackage[style = chem-rsc, citestyle = numeric-comp, sorting=none]{biblatex}
\usepackage{graphicx, geometry, hyperref, fancyhdr, titlesec, amsmath, setspace, tabularx, booktabs, array,multirow,float,tocloft, lmodern, authblk}
\usepackage[T1]{fontenc}
\usepackage[skip=10pt, indent=0pt]{parskip}
\usepackage[singlelinecheck=false]{caption}
\usepackage[dvipsnames,svgnames,x11names]{xcolor}
\usepackage{graphicx}
\definecolor{tunablue}{RGB}{50, 70, 150}
\definecolor{tunagrey}{RGB}{247,247,247}
\definecolor{tunared}{RGB}{180,0,0}
\usepackage[utf8]{inputenc}
\titleformat{\section}
  {\normalfont\large\bfseries}
  {\thesection}{1em}{}
\usepackage[most]{tcolorbox}
\newtcolorbox{mybox}{
  colback=white,
  colframe=black,
  arc=3mm,
  boxrule=0.3mm,
  width=\textwidth,
  left=3mm,
  right=2mm,
  top=2mm,
  bottom=2mm,
  fonttitle=\bfseries,
  fontupper=\ttfamily\small\color{black}
}

\newcommand{\tunabox}[4]{
\begin{mybox}
  \color{tunablue}TUNA\color{black}~#1~\color{tunared}:\color{black}~#2
  \color{tunared}:\color{black}~#3 
  \color{tunared}:\color{black}~#4
\end{mybox}
}

\usepackage[citestyle=numeric-comp, sorting=none]{biblatex}
\newgeometry{top=3cm, bottom=3cm, outer=3cm, inner=3cm}
\hypersetup{colorlinks=true, linkcolor=black, urlcolor=black, pdftitle={TUNA: A streamlined quantum chemistry program for atoms and diatomics}, citecolor=tunablue}
\title{\huge 
TUNA: A streamlined quantum chemistry\\ program for atoms and diatomics}
\date{}
\author{Harry Brough$^*$}
\affil{\small \it Department of Chemistry, The University of Manchester,
Manchester, M1 7DN, United Kingdom}
\affil{$^*$ harry.brough@manchester.ac.uk}

\begin{document}

\maketitle

\section*{Abstract}

We present TUNA, an open-source quantum chemistry program specifically designed for atoms and diatomic molecules. Within this narrow molecular domain, a broad and consistent set of electronic structure methods and calculation types is available. Energies, optimisations, vibrational frequencies, response properties, coordinate scans and \textit{ab initio} molecular dynamics trajectories can all be accessed through an intuitive command-line interface.

A single principle underlies TUNA: once a method can be used to evaluate the energy, all properties follow from numerical differentiation. This makes the program both a transparent teaching platform and a compact environment for benchmarking methods on diatomic molecules --- among the most simple yet instructive systems in quantum chemistry. Reference implementations including density functional theory, many-body perturbation theory and coupled cluster theory, supported by detailed theoretical documentation, make TUNA an accessible foundation for developing improved methods and algorithms in electronic structure.

\vfill

\begin{figure}[!b]
    \centering
    \includegraphics[width=0.52\linewidth]{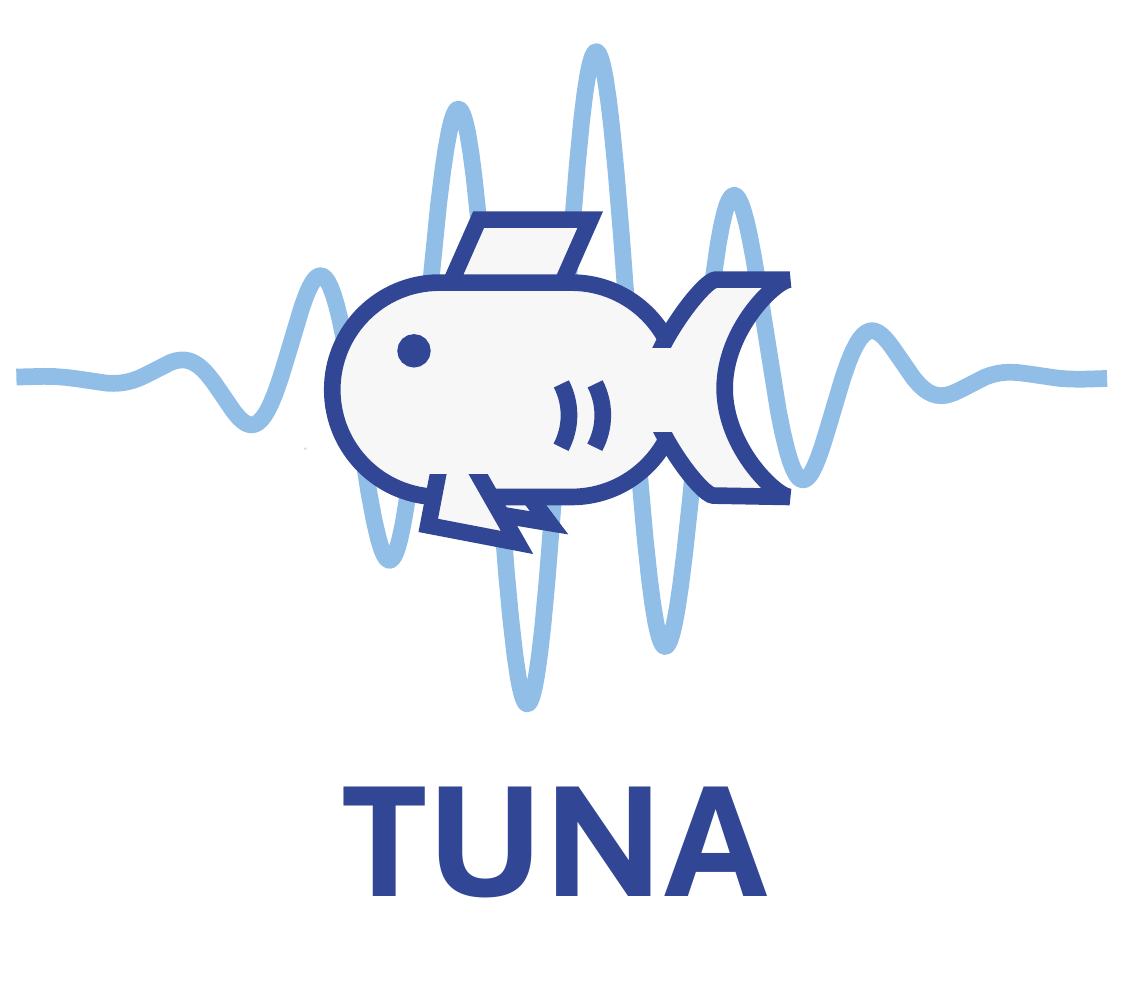}
\end{figure}

\section{Introduction}\label{intro}

Quantum chemistry programs are typically designed for broad applicability. Such breadth is valuable, but also brings complexity: inputs become longer, workflows are method-dependent, and not all properties are available at all levels of theory. TUNA takes the opposite approach. 

It is designed specifically for atoms and diatomics, with a compact command-line syntax of 

\begin{center}
    \tunabox{[Calculation]}{[Atom A] [Atom B] [Distance]}{[Method] [Basis]}{}
\end{center}

plus optional keywords, for the calculation types listed in Table \ref{tab:calculations}. For example a restricted Hartree--Fock optimisation of the hydrogen molecule in the 6-31G basis set is simply

\begin{center}
    \tunabox{OPT}{H H 1.0}{HF 6-31G}{}
\end{center}

This uniformity makes comparison between electronic structure methods straightforward and ensures TUNA is easily accessible to non-expert users. 

Diatomic molecules are among the most instructive systems in quantum chemistry. They are simple enough to analyse along a single bond coordinate, yet rich enough to expose the full range of challenges in electronic structure theory: bond breaking, static correlation, spin symmetry breaking, spectroscopy, and excited state behaviour. They can therefore serve equally well as teaching systems and research benchmarks. 

By the symmetry of diatomics, many observables reduce to one or two independent components, which makes numerical differentiation practical for many molecular properties. TUNA exploits this symmetry throughout the program, handling response properties \textit{via} optimised finite difference schemes and computing anharmonic vibrational frequencies directly on the one-dimensional potential energy surface.

\begin{table*}[b!]
    \centering
    \caption{Calculation types implemented in TUNA and their respective keywords}
    \begin{tabularx}{\textwidth}{@{} l X @{}}
        \toprule
        Keyword & \hspace{20mm}Calculation Type\\
        \midrule
        \texttt{SPE}    & \hspace{20mm}Single point energy\\
        \texttt{SCAN}    & \hspace{20mm}Nuclear coordinate scan\\
        \texttt{OPT}    & \hspace{20mm}Geometry optimisation\\
        \texttt{FORCE}    & \hspace{20mm}Force on nuclei\\
        \texttt{BDE}    & \hspace{20mm}Bond dissociation energy\\
        \texttt{IP}    & \hspace{20mm}Ionisation potential\\
        \texttt{EA}    & \hspace{20mm}Electron affinity\\
        \texttt{FREQ}    & \hspace{20mm}Harmonic frequency\\
        \texttt{OPTFREQ}    & \hspace{20mm}Optimisation and subsequent harmonic frequency\\
        \texttt{ANHARM}    & \hspace{20mm}Non-perturbative anharmonic frequency\\
        \texttt{MD}    & \hspace{20mm}\textit{Ab initio} molecular dynamics\\
        \bottomrule
    \end{tabularx}
    \label{tab:calculations}
\end{table*}

\begin{figure}[t]

    \centering
    
    \includegraphics[width=\textwidth]{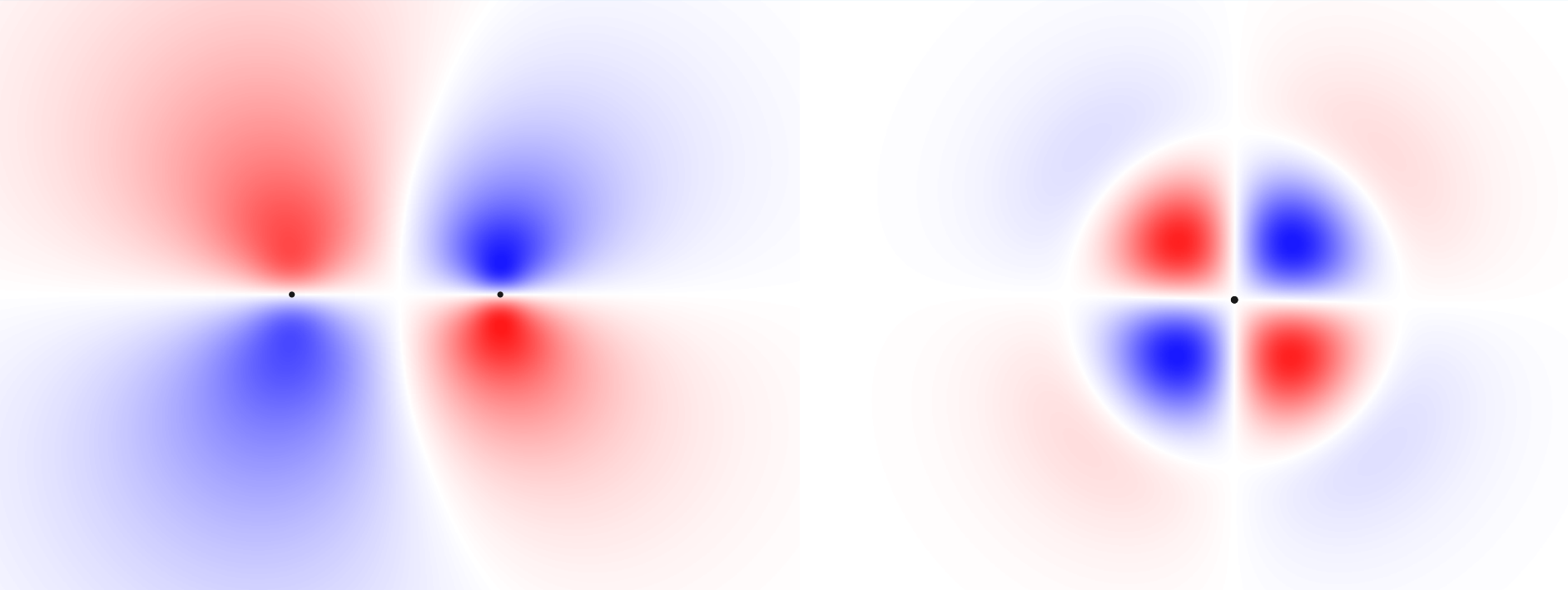}

    \caption{Molecular orbital plots calculated in TUNA. The HF/6-31G $\pi^*$ LUMO of carbon monoxide (left) and a 4d virtual orbital from PBE/cc-pVQZ on the hydrogen atom (right). The orbital phases are shown in red and blue, separated by white nodes. The symmetry of diatomic molecules means that molecular orbitals can be fully described in two dimensions.}
    
    \label{fig:orbitals}

\end{figure}

The TUNA software is well suited to teaching and exploratory work because it exposes modern electronic structure methods through a transparent interface. A user can begin with a restricted Hartree--Fock calculation on the hydrogen molecule, examine orbital energies, charges, and bond orders, then move directly to unrestricted references, perturbation theory, coupled cluster, and density functional methods. Because the systems are small, changes in method can be interpreted directly and without ambiguity. Intermediate data structures and algorithmic pathways are fully accessible, and a built-in interface to Matplotlib \cite{Matplotlib} supports plotting of molecular orbitals, electron densities, and potential energy curves through keywords such as \texttt{PLOTHOMO}, \texttt{PLOTLUMO}, and \texttt{DENSPLOT}, shown in Figure \ref{fig:orbitals}. The TUNA manual accompanies the program with a fully referenced theoretical background to each method, as well as dozens of ``simple input line'' examples which can be copied and pasted into a terminal.

For researchers, TUNA is a practical development platform. Its narrow molecular scope means new methods can be implemented, tested, and debugged quickly, with results that are directly interpretable. The program is efficient and flexible enough for meaningful benchmarking on diatomics, and its thoroughly documented Python codebase makes modifications straightforward. TUNA therefore addresses three audiences at once: students learning electronic structure theory, researchers benchmarking methods on small molecules, and developers prototyping new algorithms.

\section{A common framework for response properties}
The symmetry of diatomics makes numerical differentiation practical as a general strategy rather than a last resort. Within TUNA, structural, vibrational, and electric field response calculations are handled within this same framework. Optimised structures and harmonic frequencies are obtained from numerical first and second derivatives respectively, while finite field methods provide access to dipole moments, polarisabilities, and hyperpolarisabilities.

For example, the energy gradient at bond length $R$ is calculated by central differences as

\begin{equation}
    \frac{\partial E}{\partial R} \approx \frac{E(R+h)-E(R-h)}{2h}\;.
\end{equation}

Second derivatives are calculated similarly with a five-point stencil, third derivatives with an eight-point stencil and fourth derivatives with a nine-point stencil. The parameter $h$ in all cases has been optimised to maximise precision, and tightened energy convergence criteria are used in calculations requiring these higher derivatives. Once an electronic structure method is implemented, our numerical approach means it can immediately be used throughout the whole program --- to calculate structures, molecular dynamics trajectories, and response properties.

The complete basis set limit can be estimated by combining energies from two calculations with successively larger basis sets \cite{Neese2011}. Since observables in TUNA are numerical derivatives of the energy, approximate complete basis response properties become automatically available. 

This approach can provide experimental accuracy. For example, running the command
\begin{center}
    \tunabox{BDE}{H H 1.0}{CISD cc-pVQZ}{ZPE VPT2 EXTRAPOLATE}
\end{center}
instructs TUNA to calculate the full configuration interaction bond dissociation energy of the hydrogen molecule, with basis set extrapolation and correction for zero-point energy with anharmonic effects from second-order vibrational perturbation theory, \texttt{VPT2} \cite{Piccardo2015}. The calculated bond dissociation energy of 103.32~kcal~mol\textsuperscript{$-$1} compares to the experimental value of 103.27~kcal~mol\textsuperscript{$-$1} \cite{Liu2009}, well within ``chemical accuracy'' of 1~kcal~mol\textsuperscript{$-$1}. Importantly, this quantitative level of accuracy is obtained through a single concise command.
\section{Calculation types}\label{sec:calculations}

\begin{figure}[t]

    \centering
    \includegraphics[width=\linewidth]{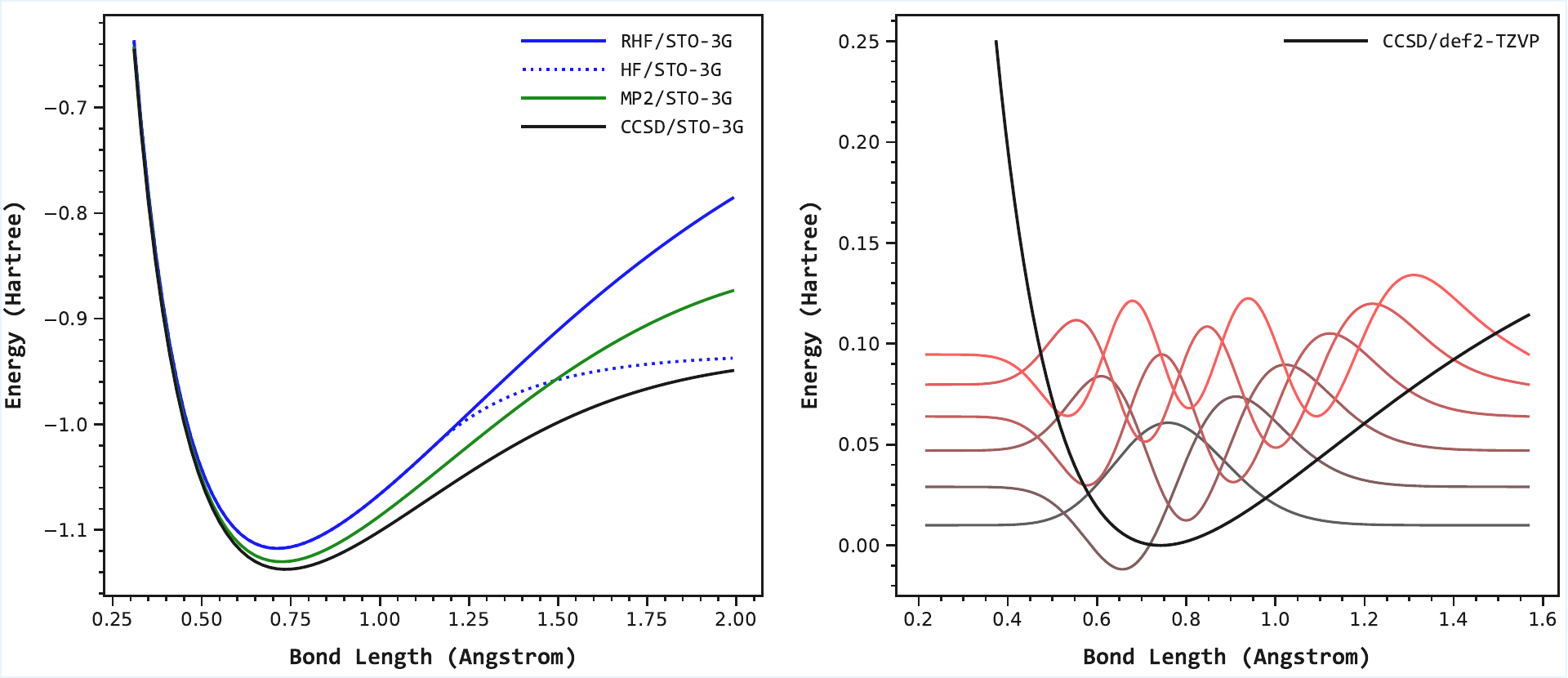}
    \caption{The dihydrogen potential energy surface calculated by electronic structure methods in the STO-3G basis set with a \texttt{SCAN} calculation in TUNA, shown with the \texttt{SCANPLOT} and \texttt{ADDPLOT} keywords (left). The anharmonic vibrational wavefunctions (increasing red intensity with energy) calculated with CCSD/def2-TZVP and shown with the \texttt{VIBPLOT} keyword (right).}
\label{fig:vibes}   
\end{figure}

\begin{figure}[t!]
    \centering
    \includegraphics[width=\textwidth]{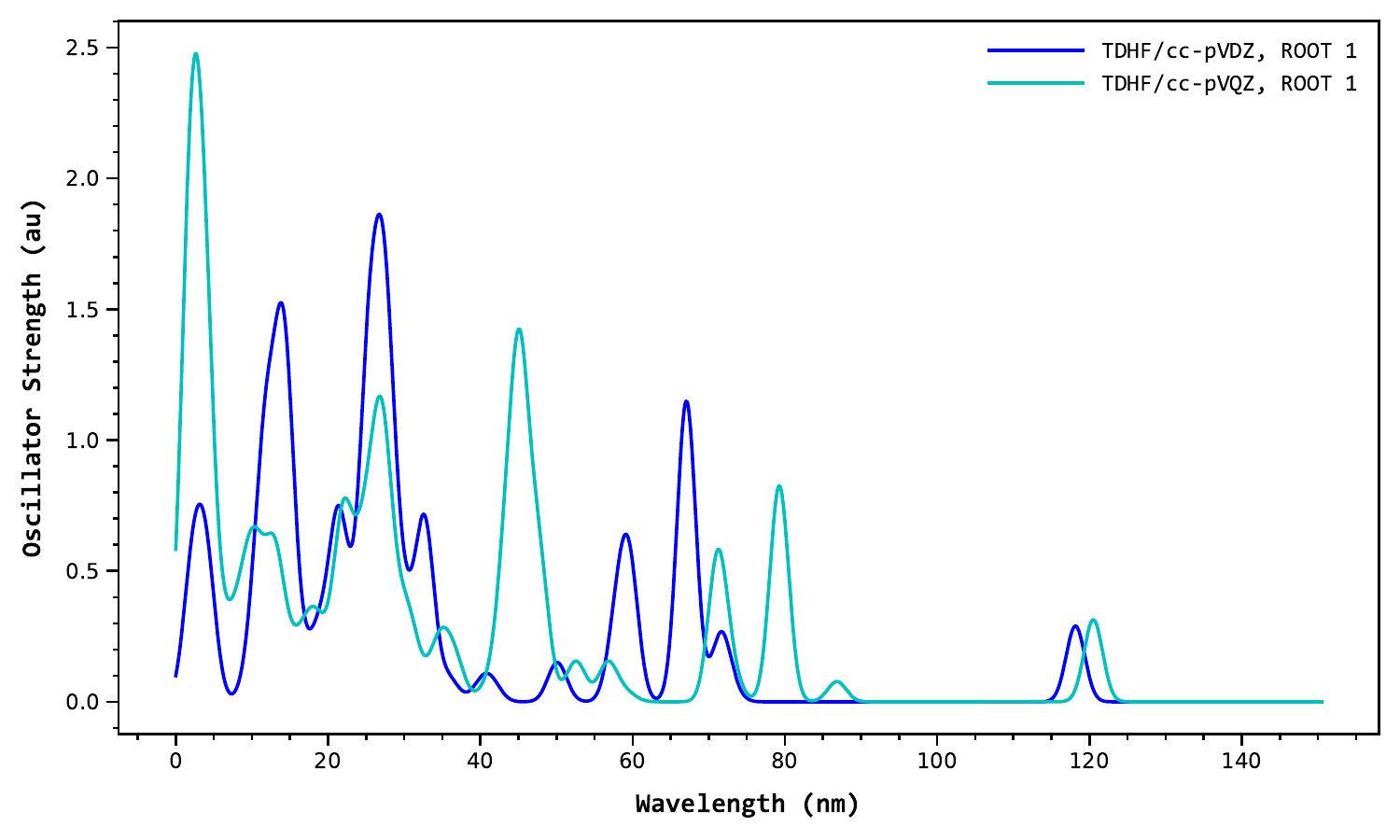}
    \caption{The absorbance spectrum of carbon monoxide in two basis sets, by an excited state \texttt{SPE} calculation in TUNA, shown with the \texttt{ABSPLOT} and \texttt{ADDPLOT} keywords.}
    \label{fig:absorbance-spectrum}
\end{figure}

Beyond single point energies, TUNA can calculate optimised geometries, vibrational frequencies and \textit{ab initio} molecular dynamics trajectories with the velocity Verlet algorithm. There are also built-in calculation type keywords to determine bond dissociation energies, ionisation potentials and electron affinities. Nuclear coordinate scans are available with a Matplotlib interface, and multiple potential energy surfaces can be overlaid on the same plot with the \texttt{ADDPLOT} keyword for comparison, with examples shown in Figure \ref{fig:vibes} and Figure \ref{fig:absorbance-spectrum}.

Geometry optimisations use Newton's method, which in one dimension reduces to a scalar step with no ambiguity in the search direction. Approximate Hessians are calculated by default to accelerate convergence, and ``exact'' Hessians can be requested. An exact Hessian is also calculated in a harmonic frequency calculation to determine the normal mode frequency by

\begin{equation}
    \omega^2 = \frac{1}{\mu}\frac{\partial^2E}{\partial R^2} 
\end{equation}

where $\mu$ is the reduced mass. This vibrational frequency is used to calculate thermochemical results including the zero-point energy, enthalpy, entropy and Gibbs free energy. Harmonic infrared intensities are also available \textit{via} numerical dipole moment derivatives.

Another calculation type determines the bond dissociation energy between atoms A and B,

\begin{equation}
    E_\text{BDE} = E(\text{A})+E(\text{B}) - E(\text{A--B})\;,
\end{equation}

which is counterpoise corrected by TUNA through the use of ``ghost'' basis functions \cite{Boys1970}. Zero-point energy correction can be requested with the \texttt{ZPE} keyword, with an optional anharmonic treatment from \texttt{VPT2}. Ionisation potentials and electron affinities are adiabatic by default, but a static geometry calculation can be requested with the \texttt{VERTICAL} keyword.

The one-dimensional potential energy surfaces of diatomic molecules mean that the nuclear Schr\"odinger equation can be solved ``exactly'' for the vibrational energy levels and wavefunctions, $\Psi(R)$,  plotted by TUNA in Figure \ref{fig:vibes}. Infrared intensities for transitions between state $i$ and $j$ can also be calculated by numerically evaluating the transition dipole moment,

\begin{equation}
    \mathcal{A}_{ij}=\int\Psi^*_{i}(R)\,\mu(R)\,\Psi_j(R)\,\text{d}R\;,
\end{equation}

where the dipole moments are derivatives of the energy with respect to a finite electric field. Both the dipole and potential energy surfaces are interpolated before diagonalisation of the nuclear Hamiltonian. A surface scan extends outwards from the equilibrium geometry and the Hamiltonian construction and diagonalisation process repeats until the fundamental frequency converges, minimising the number of energy evaluations without loss of accuracy. As with all calculation types, basis set extrapolation is available and can provide sub-cm\textsuperscript{$-$1} accuracy.

\section{Electronic structure methods and basis sets}

While narrow in molecular scope, TUNA has a diversity of basis sets and electronic structure methods, the vast majority of which are available for both spin-restricted and unrestricted references. For several methods, TUNA is one of few, if any, open source implementations.

Methods in TUNA include self-consistent Hartree theory and Hartree--Fock theory, and many-body perturbation theory methods including MP2, MP3, and MP4. Spin-component-scaled modifications and density matrix-based AO-MP2 are also available \cite{Surjan2005, Moller1934:46, Grimme2003:118, Grimme2003:24}. Self-consistent density functional approximations are implemented from all rungs of the ``Jacob's ladder'' of exchange--correlation functionals, with components listed in Table \ref{tab:functionals} \cite{Perdew2001}, including hybrid functionals like PBE0 \cite{Adamo1997} and B3LYP \cite{Stephens1994}, which incorporate a proportion of ``exact'' Fock exchange, and double-hybrid functionals like B2PLYP \cite{Grimme2006} and DSD-BLYP \cite{Kozuch2010} which also include a proportion of second-order perturbation theory correlation.

Sophisticated correlated wavefunction methods based on configuration interaction (CI) and coupled cluster (CC) theory are also available, including CCD, CISD, CCSD, CCSD(T), CISDT, CCSDT, CCSDT(Q) and CCSDTQ. The quadratic configuration interaction methods QCISD and QCISD(T) are implemented, as are approximate coupled cluster methods CC2 and CC3 and methods based on the coupled electron pair approximation, CEPA \cite{Jiang2025,Jiang2025:1,CCSDT, HAMPEL19921, RENDELL1991462}. 

Excited states and their transition oscillator strengths can be calculated with time-dependent Hartree--Fock theory or configuration interaction singles, with an optional perturbative doubles correction \cite{Head-Gordon1994:219}. A time-dependent DFT implementation is also available, with support for local spin density approximation exchange--correlation functionals, including hybrid and double-hybrid modifications. Figure \ref{fig:absorbance-spectrum} shows the use of the \texttt{ABSPLOT} keyword which enables straightforward comparison of absorbance spectra between electronic structure methods.

\begin{table*}[t!]
    \centering
    \caption{Exchange and correlation functionals implemented in TUNA \cite{Slater1951, Perdew1986, Perdew1992, Perdew1996, Becke1988, Tau2003, Vosko1980, Lee1988, Mardirossian2015, Grimme2006-27, Becke1997, Furness2020, Sun2015}}
    \begin{tabularx}{\textwidth}{@{} l @{\hspace{15mm}} X @{}}
        \toprule
        Type & Functionals\\
        \midrule
        Exchange    & Slater, (rev)PBE, RPBE, B88, mPW, PW91, (rev)TPSS, (r\textsuperscript{2})SCAN, B97(-D), B97M-V \\
        \midrule
        Correlation & VWN3, VWN5, PWLDA, P86, PW91, PBE, LYP, (rev)TPSS, (r\textsuperscript{2})SCAN, B97(-D), B97M-V\\
        \bottomrule
    \end{tabularx}
    \label{tab:functionals}
\end{table*}

A wide range of contracted Gaussian-type orbital basis sets are available. These include minimal basis sets such as STO-3G, Pople style sets such as 6-31G* \cite{hariharan1973a, frisch1984a, hehre1972a}, the Dunning style cc-pV$N$Z sets up to triply-augmented sextuple-$\zeta$ \cite{Dunning1989:90, woon1995a, peterson2002a, Woon1993:98}, the Ahlrichs' def2- basis sets \cite{weigend2005a}, Jensen's pc(seg)-$N$ sets \cite{jensen2001a, jensen2002b, jensen2002a, jensen2014a} and atomic natural orbital basis sets, ano-pV$N$Z \cite{neese2011a}.

Methods can be modified with simple keywords. For example, the proportion of exact Hartree--Fock exchange in an exchange--correlation functional is easily changed through the \texttt{HFX} keyword, the same-spin scaling parameter of SCS-MP2 can be changed with \texttt{SSS} and the amplitude convergence criteria for coupled cluster calculations is altered with \texttt{AMPCONV}. Any number of optional  keywords can be used together in the TUNA input line, in any order.

\section{Implementation}

The philosophy of TUNA mirrors that of Python: both trade raw speed for simplicity and clarity. The program is implemented in Python 3, making extensive use of NumPy \cite{Harris2020:585} and SciPy \cite{Virtanen2020:17} for numerical operations and Matplotlib \cite{Matplotlib} for integrated plotting. Vectorised NumPy tensor contractions --- particularly through the \texttt{einsum} function ---  keep the code compact and readable while supporting high-order correlated methods. Beyond these modules, dependencies have been kept to a minimum to make installation as simple as possible.

Diatomic symmetry is exploited throughout to improve efficiency, most significantly in the McMurchie--Davidson evaluation of molecular integrals over Gaussian basis functions, where the parity of a linear molecule aligned on the $z$-axis yields substantial speedups. Because these integrals are difficult to vectorise this module is implemented in Cython \cite{Cython} --- a compiled extension language for Python. Utilisation of symmetry, avoiding writing to disk and OpenMP multithreading in the Cython code makes TUNA competitive in speed with well-established general-purpose quantum chemistry codes \cite{Dagum1998}.

\section{Summary and perspectives}

This article has presented TUNA, a user-friendly electronic structure program for atoms and diatomics combining a simple command-line interface with a diversity of electronic structure methods, basis sets and calculation types. By narrowing its molecular scope, TUNA finds a clear niche as a streamlined laboratory for quantum chemistry on the smallest molecules.

Several developments are planned. Relativistic corrections are a priority for improving accuracy on heavier diatomics. Magnetic response properties \textit{via} finite magnetic fields and expanded configuration interaction capabilities, including CASSCF, are under development. Closer integration of theory and implementation in the documentation is a continuing focus.

The TUNA code and documentation are available at \href{https://github.com/h-brough/TUNA}{\texttt{github.com/h-brough/TUNA}} under the MIT license and the program is easily installed from PyPI with \texttt{pip install quantumtuna}. Contributions, feature requests and bug reports through the GitHub issue tracker are welcome.

\section{Acknowledgements}

Several open-source codes and tutorials provided vital inspiration for TUNA, including the Psi4NumPy reference implementations and tutorials \cite{Smith2018}, the HarPy suite of Hartree--Fock-based electronic structure codes \cite{harpy}, the Crawford group programming projects \cite{crawford-tuts}, and the PyDFT program with its excellent textbook-style documentation \cite{filot2025electronic}. The ORCA quantum chemistry package was also a key inspiration for its ``simple input line'' approach \cite{Neese2012:2, Neese2022:12}. 

The author is also very grateful to Hannah Whittome for her brutal criticism of an early version of the TUNA logo, which turned out to be entirely justified.

\printbibliography[title=References]
\end{document}